# RESPONSE TO RFI "SCIENCE OBJECTIVES AND REQUIREMENTS FOR THE NEXT NASA UV/VISIBLE ASTROPHYSICS MISSION CONCEPTS"

## MASSIVE STARS: KEY TO SOLVING THE COSMIC PUZZLE

Wofford Aida[1], wofford@stsci.edu, 410-338-4450; Leitherer Claus[1], leitherer@stsci.edu; Walborn Nolan R.[1], walborn@stsci.edu;
Smith Myron[1], msmith@stsci.edu; Peña-Guerrero María A.[1], pena@stsci.edu;
Bianchi Luciana[2], bianchi@pha.jhu.edu; Thilker David[2], dthilker@pha.jhu.edu; Hillier D. John[3], hillier@pitt.edu;
Maíz Apellániz Jesús[4], jmaiz@iaa.es; García Miriam[5], mgg@iac.es; Herrero Artemio[5], ahd@iac.es;
[1]Space Telescope Science Institute; [2]Johns Hopkins University; [3]University of Pittsburgh;
[4]Instituto de Astrofísica de Andalucía-CSIC; [5]Instituto de Astrofisica de Canarias and Universidad de La Laguna

**Abstract.** We describe observations in the nearby universe (<100 Mpc) with a ≥10-m space-based telescope having imaging and spectral capabilities in the range 912-9000 Å that would enable advances in the fields of massive stars, young populations, and star-forming galaxies, that are essential for achieving the Cosmic Origins Program objectives i) <u>how are the chemical elements distributed in galaxies and dispersed in the circumgalactic and intergalactic medium; and ii) when did the first stars in the universe form, and how did they influence their environments</u>. We stress the importance of observing hundreds of massive stars and their descendants individually, which will make it possible to separate the many competing factors that influence the observed properties of these systems (mass, composition, convection, mass-loss, rotation rate, binarity, magnetic fields, and cluster mass).

## 1. ESSENTIAL ADVANCES IN MASSIVE-STAR ASTROPHYSICS

**Motivation.** Massive stars (≥8 $M_\odot$) play seminal roles in the chemical enrichment and evolution of galaxies. They produce the bulk of the α-elements, e.g., oxygen (Weaver & Woosley 1995, Woosley & Weaver 1995), some of the Fe and Fe-peak elements (François et al. 2004), and at low metallicity, significant amounts of C and N (Pettini et al. 2008, Chiappini et al. 2006). They also craft the interstellar medium through their winds and explosive deaths (Mac Low & McCray 1988), enrich the circumgalactic medium (CGM, Tumlinson et al. 2011), dominate the spectral energy distribution of star-forming galaxies (Leitherer et al. 1999), are essential for understanding the energetics of galactic centers, and are related to very energetic and disruptive processes like supernovae (II, Ib/Ic, and Ia) and possibly long gamma-ray bursts (Dessart et al. 2012). In addition they produce (Gall et al. 2011) and destroy dust. Finally, massive stars may be responsible for causing reionization in the early Universe (Robertson et al., 2010; Heckman et al. 2011). Interesting physics of massive stars include atomic processes, radiative transfer in complex media, stellar wind/ISM interactions, interacting binary evolution, radiation-(magneto)hydrodynamics, shocks, and production of X-rays.

**The winds of blue massive stars (BMSs).** The term blue massive stars encloses OB stars with masses ≥20 $M_\odot$, Wolf-Rayet (WR) stars, and Luminous Blue Variables (LBVs). These stars experience radiation-driven outflows of matter, or winds, propelled by the absorption of photons in the numerous UV transitions of metallic ions (Kudritzki & Puls, 2000). The stellar wind is the main interface of pre-supernova massive stars with the interstellar medium through the injection of mechanical and radiative energy, and chemically enriched material. The wind is also a principal agent of the evolution of the massive star itself. Because of the wind mass removal, the rate and efficiency of the central nuclear reactions change, altering the duration of the evolutionary stages and ultimately deciding the fate of the star (the SN explosion, the stellar yields, and the remnant; Georgy et al., 2009; Matteucci, 2008; Woosley et al., 2002). The stellar wind leaves a significant imprint on the observed spectra of BMSs at different wavelength ranges (see the review by Kudritzki & Puls 2000 and more recently Martins 2011). Its main parameters, mass loss rate ($\dot{M}$) and terminal velocity ($v_\infty$) can consequently be derived from quantitative spectral analysis using model atmospheres that include radiation driven winds (e.g., Puls et al., 2005; Pauldrach et al., 2001; Hillier & Miller, 1998; Bianchi et al., 2009). However, the wind signatures also alter line diagnostics for other stellar parameters. For instance, the wind enhances the line blanketing effect and fills the He II lines, which are critical to determine $T_{eff}$ in O stars (Repolust et al., 2004). The photospheric and wind parameters must be determined jointly.

**Uncertain mass-loss rates.** In order to interpret distant unresolved starbursts and their effects on their environments, but also feedback into the ISM and CGM, it is essential to understand the structure and evolution of massive stars observable at high spectral and spatial resolutions. That is far from the case at present, even for their relatively quiescent early evolutionary phases, and it is partly due to remaining unexplained points regarding the winds of BMSs. For instance, the issues of inhomogeneities and asymmetries in their powerful winds, which substantially affect their mass-loss rate estimates, and therefore, our knowledge of their evolution, are matters of debate. In addition, the potential dependences of these effects on metallicity have hardly been addressed at all.

**Unknowns in the final stages of massive stars.** The situation regarding our knowledge of the advanced evolution and endpoints of massive stars is even worse: there is a zoo of peculiar objects of uncertain or unknown interpretation, and the progenitors of the diverse categories of core-collapse supernovae are unidentified except for those of the lowest masses. The foregoing refers primarily to simple, single stars; complicating factors such as rapid rotation, magnetic fields, and binary interactions are only beginning to be addressed. Vast observational and theoretical efforts will be required to improve this situation.

**Ultraviolet is key.** The far-ultraviolet (900-1200 Å, FUV) and ultraviolet (1200-2000 Å, UV) are the only ranges where terminal velocities of O and early-B stars can be derived from. The value of $v_\infty$ is calculated from the bluest extent of the Doppler-shifted absorption of P-Cygni profiles of resonance lines in these spectral ranges: O VI 1031.9,1037.6, S IV 1062.7,1073.0,1073.5, the C IV 1169+C III 1176 blend, N V 1238.8,1242.8, Si IV 1393.8,1402.8 and C IV 1548.2,1550.8 (Kudritzki & Puls, 2000; Martins, 2011). When no such spectra are available, the terminal velocity of OB stars is estimated from calibrations with spectral type (Kudritzki & Puls, 2000) or from the escape velocity ($v_\infty/v_{esc} \approx 2.6$ for $T_{eff} > 21,000$K; $v_\infty/v_{esc} \approx 1.3$ for $T_{eff} < 21,000$K, Lamers et al. 1995), and then scaled with metallicity ($v_\infty \propto Z^{0.13}$, Leitherer et al. 1992). However we need to produce calibrations for critical transitional phases (early WR, Ofpe/WN9 stars etc.), where $\dot{M}$ varies highly. The UV lines are also essential (combined with e.g., H$\alpha$) to constrain clumping factors, which are needed for a precise measurement of $\dot{M}$ (Bouret et al. 2005; Fullerton et al. 2006; Herald & Bianchi 2011; Bianchi et al. 2009).

**Metallicity and wind parameters.** Theory predicts a strong correlation between the momentum carried by the wind and the luminosity of the star and metallicity: the wind-momentum luminosity relation (Kudritzki et al., 1995). This relation has been proved empirically, and its metallicity dependence empirically characterized from the Milky Way (MW) down to the metallicity of the Small Magellanic Cloud (SMC) (Vink et al., 2001; Mokiem et al. 2007). Very low metallicity BMSs are expected to experience weaker winds than SMC stars and much weaker winds than MW stars (see Fig. 1). However, some recent results are in marked contrast. Tramper et al. (2011) reported 6 stars with stronger wind momentum than expected at the poor metallicity of their host galaxies (~ 1/7 $Z_\odot$, IC1613, WLM and NGC3109) from X-Shooter spectroscopic analyses, although error bars are too large for results to be determinant. Herrero et al. (2012) report the analysis of an Of star in IC1613 that may also have a strong wind or, alternatively, a lower than expected wind acceleration. These examples, if confirmed by a large sample of objects, pose a challenge to the theory, as there are few metals to drive the wind. Yet, this might explain why long-GRBs (typically associated with type Ic SN, Woosley & Heger, 2006) are mostly found in metal-poor environments (Modjaz et al., 2008; Levesque et al., 2010) but require a strong wind to remove the H and He envelope in the pre-SN stages. However, the studies of the winds of BMSs beyond the Magellanic Clouds are based only on optical data, hence lack direct measurements of the terminal velocity and introduce large error bars in the wind-momentum luminosity relation.

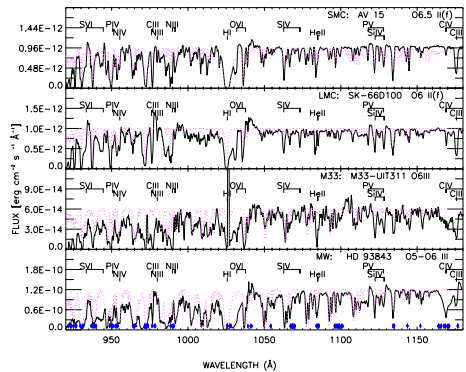

**Fig. 1** FUSE spectra of BMSs with approximately the same spectral type in the SMC, LMC, outer-M33 and MW. Diamonds mark interstellar and airglow transitions, while the dotted line is a model for hydrogen absorption towards the targets. The stellar spectra (black solid line) display lines of OVI, PV and CIII which develop a weaker wind profile (even photospheric) as the metallicity decreases, i.e., as we move upwards in the plot.

**Past UV observations.** The *International Ultraviolet Explorer* (*IUE*) provided a sample of ~200 Galactic OB spectra at high resolution ($R\sim10^4$) in the 1200-1900 Å range. The *Far Ultraviolet Spectroscopic Explorer* (*FUSE*) observed a few tenths of stars from 900-1200 Å in both the Galaxy (Pellerin et al. 2002) and the Magellanic Clouds (MCs, Walborn et al. 2002), the latter providing vital information at lower metallicities (about 1/2 solar in the Large Magellanic Cloud and 1/5 in the Small Magellanic Cloud). However, the *Hubble Space Telescope* has so far failed to realize its potential to provide the corresponding 1200-1900 Å data for an adequate sample in the Magellanic Clouds, only a few tens of OB UV spectra having been observed at high resolution in the Small Cloud, and even fewer in the LMC. The existing *HST* sample does not cover the relevant parameter space either for OB astrophysics or as a reference to model distant starbursts (lower metallicities than in the MCs are required). The main reasons for this shortcoming appear to be the heavy oversubscription of *HST* by multiple instrument configurations and the inefficiency of observing individual stars.

**Role of a ≥10-m telescope.** i) UV spectroscopy for a statistically significant sample of OB stars in the Local Group galaxies, including the Magellanic Clouds. A multiple-object UV spectrograph would enable to obtain an adequate sample, just as large ground-based telescopes have in the optical. For instance, the Sanduleak objective-prism survey of the LMC (Cerro Tololo Inter-American Observatory Contribution No. 89, 1970) identified the 1200 brightest, isolated OB supergiants, including the progenitor of SN~1987A, which was unfortunately not seriously observed even in the optical prior to the arrival of the event. One can estimate that there are 1000 SN events on the way from the MCs; hopefully we might have done better before the next one arrives. If the spectrograph could do 100 objects at a time, this survey would be covered in 12 exposures; of course, multiple exposures are desirable to improve signal-to-noise and address variability. However, this sample is the tip of an iceberg. For instance, the (ESO) VLT-FLAMES Tarantula Survey in 30~Doradus/LMC (Evans et al. 2011), the largest starburst region in the Local Group, has obtained high-resolution spectroscopy of 800 OB stars in that field alone, which are currently being analyzed with all state-of-the-art observational and theoretical techniques and will provide unprecedented information about massive stellar and cluster evolution on this scale. However, without the UV to constrain wind and photospheric parameters (including the bolometric luminosity and masses), the huge potential of this surveys, and of similar large investements of ground-based telescopes, cannot be fully realized. Most of these stars are fainter and more heavily extincted than those in the Sanduleak survey, but feasible with a 10-meter space telescope. We shall never understand distant starbursts until we understand the intricate, multiple stellar populations and generations in 30~Doradus, as well as other nearby objects such as Henize N11 in the LMC, a once and future 30~Doradus about 2~Myr older, which is a significant difference on the evolutionary timescales of the most massive stars. GALEX has just finished observing the entire LMC and SMC in the NUV, allowing generation of a panchromatic catalog of UV-bright stars (Thilker et al. in prep). This survey will enable careful vetting of potential targets for UV spectroscopy, thereby maximizing the return from a systematic 10-m observing campaign.

ii) Stellar astrophysics in a range of metallicity environments and conditions. The telescope would produce reliable measurements of the wind parameters, and enable firm conclusions regarding the effect of metallicity on the wind momentum of BMSs. To widen the studied metallicity range we need to reach farther into the Local Group: M31, M33 and the dwarf irregulars (Garcia et al., 2011b, Bianchi et al. 2011, 2012a, b). However, single-object UV spectroscopy beyond the MCs is very expensive in observing time, even with HST's lowest resolution spectrograph (G140L, R ~ 2600 at 1550 Å). A recent HST treasury program has imaged six nearby star-forming galaxies with a key set of filters including 2 filters shortward of the U band. Results showed a huge variety of properties of the massive star content. At a distance of 18 Mpc, I Zw 18, the most metal-poor star-forming galaxy found in the local universe is beyond the reach of current instrumentation to perform studies of stars equivalent to those done in our Galaxy and the Magellanic Clouds. For reasons still not understood, this galaxy has maintained its near-pristine chemical composition of ~1/30 the solar value ($Z_\odot \approx 0.13$) since the epoch of galaxy formation. Its stars are the closest local counterparts in terms of composition to the first stellar generation formed in the early universe. A 20 m telescope at 1500 Å will resolve 0.1 pc in I Zw 18; comparable to resolution in 30 Doradus in V from ground (see Fig. 2).

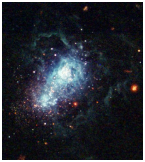

**Fig. 2.** I Zw 18. A few numbers for this galaxy are:
- D = 18 Mpc (m – M = 31.3); 88.5 pc/arcsec; $v_{helio}$ = 750 km/sec
- 370× more distant than 30 Doradus
- Individual O star has ~ $10^{-18}$ erg/s/cm$^2$/Å at 1500 Å; this is still well above the sky background
- B0 main-sequence star has V ≈ 27; solar-type star has V ≈ 36
- Foreground reddening E(B – V) = 0.03; negligible internal reddening (ideal for UV)

## 2. RESOLVED YOUNG POPULATIONS IN EXTREME ENVIRONMENTS

**Motivation.** Spatially resolved imaging and spectroscopy of young populations in uncharted extreme environments are essential for understanding hypergiant stars, and the properties of systems more massive than 30 Dor, the mini-starburst in the LMC, i.e., their star formation rates, their star formation histories, their structure, and the effects of very low metallicity on their UV spectra and their initial mass functions (IMFs).

**Hypergiant stars.** Candidates for isolated BAF-type hypergiant stars are found in the Antennae (Whitmore et al., 2010). Can they be confirmed, and what are their properties? **Higher stellar population mass regimes.** The detailed structure of starbursts several times larger and more massive than 30~Dor can also be discerned in *HST* images of the Antennae major merger at about 20 Mpc. What are their properties? Are they related to globular clusters? **Star formation rates.** In general, the infrared through Hα to UV represents a sequence of increasing age for young populations, so all three regimes are essential for accurate determination of star-formation rates. For example, NGC 604 in M33 (Fariña et al. 2012) provides a more massive version of N11 containing at least four stellar generations. **Star formation histories.** Wide-field observations of star-forming regions are important for understanding star formation histories. On small spatial scales, the OB clusters in the circumnuclear starburst region of M83 appear to have formed in a spatially uncorrelated manner, while on larger spatial scales, they show an age gradient along the starburst's arc, which is a few hundred parsecs long (Wofford et al. 2011). **The structure of superstar clusters.** NGC~5471 in M101 (García-Benito et al. 2011) is a spectacular cluster of superclusters. Improved observation and analysis of such systems will surely be relevant to more massive starbursts than 30~Dor. **Star cluster properties at very low metallcity.** Chemical composition is the key parameter that determines properties such as the stellar initial mass function (arguably), the stellar mass-luminosity relation, stellar lifetimes, the escape of stellar radiation, or release of matter from massive stars to the interstellar and intergalactic medium. Investigating the influence of chemistry on the physics of stars and resolved stellar population is a premier goal of contemporary astrophysics. As an example, I Zw 18 has a record-setting chemical composition that causes truly transformative changes in stellar properties, as opposed to incremental differences when moving, e.g., from our Galaxy to the Small Magellanic Clouds (see fig. 3). I Zw 18 will turn out to be the Holy Grail of extragalactic stellar astrophysics of massive stars.

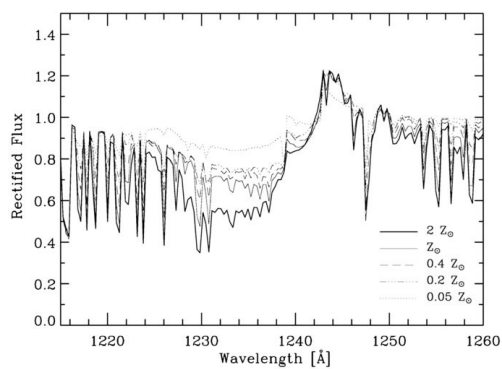

**Role of a ≥10-m telescope.** Such a telescope could provide complete CMDs of star-forming regions down 1 solar mass, IMFs, high-resolution UV/optical spectra of individual OB stars, low-resolution SEDs of individual stars, and emission- and absorption spectra of the ISM in galaxies covering a range of conditions, which are all useful for input in stellar population synthesis codes and the interpretation of unresolved young populations.

**Fig. 3** The N V P-Cygni profile of a synthetic stellar population as a function of metallicity. The metallicities of the LMC, SMC, and I Zw 18 are the bottom three values.

### 3. EXTRAGALACTIC EXTINCTION

**Motivation.** The UV-spectra of individual massive stars is very sensitive to extinction by dust and localized dust characteristics must be derived concurrently with the stellar parameters. Regarding young populations, twenty percent of their luminosity is emitted in the wavelength range 912-1200 Å, where the reddening curve peaks and where our knowledge of the reddening due to dust is still fragmentary (Leitherer et al. 2002, Buat et al. 2002, Wofford et al. in prep.). The spectral energy distribution (SED) of star-forming galaxies in the FUV is determined by the stellar IMF, the recent star formation history, and the dust extinction. Therefore, FUV observations of star-forming galaxies are fundamental for understanding the above properties and to interpret the spectra of high-z galaxies.

**Deriving selective extinction curves.** The standard technique to derive the extinction curve in stars is to observe stellar pairs of identical spectral types, one member of the pair being heavily reddened and the other unreddened. Comparison of the two SEDs allows the determination of the selective extinction curve Aλ/E(B-V). Applied to galaxies, the approach is similar, except that dust-free synthetic stellar population models are used instead of the unreddened spectrum. Alternatively, one can use low extinction observations as the unreddened galaxy spectrum.

**Past Work.** The mean UV extinction law for the SMC is usually taken as a template for low-metallicity galaxies. However, its current derivation is based on only five stars, which renders its universality questionable (Maíz Apellániz & Rubio 2012). Our scarce knowledge of extragalactic extinction has limited us to adopt average extragalactic extinction curves for heterogeneous sets of galaxies, dust compositions, and dust geometries. As illustrated in Fig. 4, it is important to push the limit with a ≥10-m telescope and produce reliable selective extinction curves.

**Role of a ≥10-m telescope.** A wide-field imager with a slitless spectrograph would enable an in-depth study of the dust extinction in the MCs and beyond by producing i) low-resolution spectra and ii) panchromatic photometry of resolved stars that would characterize their SEDs. The range 912-3000 Å is key as it includes the FUV and the 2175 Å extinction bump, which yields information about the dust composition. *HST* has been able to detect individual bright stars in the Local Group galaxies and a larger telescope will be able to probe fainter.

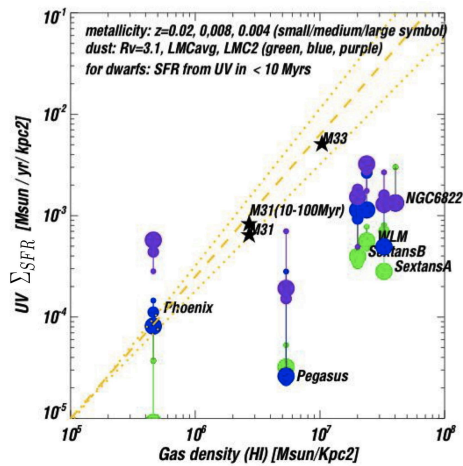

**Fig. 4** The importance of proper reddening corrections to derive star-formations rates from UV fluxes. Average SFR per unit area versus gas density for 6 Local Group dwarfs, derived from GALEX measurements of individual star-forming sites in these galaxies, assuming 3 different extinction curves, and E(B-V) derived for each region from the massive stars it contains (characterized with HST data). The figure (Bianchi et al. 2011) illustrates how the results depend on metallicity and assumed dust extinction curve. Yellow lines show the "Kennicut-Schmidt law" which was defined for disk galaxies (Rowchowdhury et al. 2009).

## 4. Lyα FROM GALAXIES AT z~0

**Motivation.** Lyα is the only emission line that we can detect from the highest redshift galaxies (z>6). Thus, it is our only probe of the internal structure of these galaxies, and it is one of our few diagnostic tools for studying the IGM at these high redshifts. The advent of WFC3 on HST has resulted in the detection of galaxies out to z ~ 10. If at these high-redshifts we can determine the detailed Lyα line profile that emerges from the galaxy, and the fraction of the line flux that escapes the galaxy, then we can use Lyα as a probe of the ionization history of the intergalactic gas, and study the nature and evolution of high-redshift galaxies. In order to determine the intrinsic Lyα properties of distant galaxies, it is useful to study lower-redshift samples where we can separate the effect of the IGM from the galaxy. In addition, detailed studies of the Lyα escape problem are only possible at low z, where the galaxies are resolved.

**Results from past-studies.** Low-redshift HST/ACS/SBC Lyα-line images have uncovered the presence of a diffuse Lyα component around concentrated knots of star-formation (Ostlin et al. 2009) that contributes to a significant fraction of the total Lyα emission from the galaxies. These images and recent HST/COS spectroscopy (Wofford et al., subm.) show the simultaneous presence of Lyα in emission and absorption in the same object within spatial scales of a few parsec, which highlights the importance of the ISM geometry in determining the escape fraction of Lyα photons from galaxies.

**Role of a ≥10-m telescope.** A wide filed telescope with imaging and spectroscopic capabilities, and having a spectral resolution of FWHM ≤ 50 km/s in the far-ultraviolet would enable a) to observe a statistically significant sample of nearby isolated disk galaxies with different inclinations so that the Lyα escape fraction can be studied as a function of inclination (unprecedented and essential); b) create maps of the Lyα emission and dust distribution in galaxies down to the spatial scales of ISM and circumgalactic clumps, i.e., <1 pc, using appropriate extinction curves (see previous section, unprecedented), and c) understand the relative importance of dust, H I column density, gas outflows, gas geometry, and stellar-population properties on the escape fraction and line profile of Lyα photons, at low-redshift.